\documentclass[aps,prd,showpacs,twocolumn]{revtex4}

\begin{document}

%\preprint{CSSM}

%\begin{flushleft}
%%hep-ph/0508078 \\
%ADP-03-119/T557 \\
%%May 2003
%\end{flushleft}

\title{Properties of the Charmed P-wave Mesons}
 \author{Stephen Godfrey\footnote{Email address: 
godfrey@physics.carleton.ca}}
\affiliation{
Ottawa-Carleton Institute for Physics \\
Department of Physics, Carleton University, Ottawa,  Canada K1S 5B6 }
%\maketitle

\date{\today}

\begin{abstract}
Two broad charmed mesons,  the $D_0^*$ and $D_1'$,
have recently been observed.  
We examine the quark model predictions for the $D_0^*$ and $D_1'$ 
properties and 
discuss experimental measurements that can shed light on them. 
We find that these states are well described 
as the broad, $j=1/2$ non-strange charmed $P$-wave mesons.
Understanding the $D_0^*$ and $D_1'$ states can provide 
important insights into the $D_{sJ}^*(2317)$, $D_{sJ}(2460)$ states 
whose unexpected properties have led to 
renewed interest in hadron spectroscopy.  
\end{abstract}

\maketitle

\section{Introduction}

It has long been pointed out that light-heavy mesons act as the
hydrogenic atoms of QCD and represent a unique laboratory to test our 
understanding of QCD \cite{DeRujula:1976kk,Rosner:1985dx}.  
In the limit that the heavy quark mass becomes infinite,
the properties of the meson are determined by those of the light 
quark \cite{DeRujula:1976kk,Rosner:1985dx,isgur91}. 
The light quarks are characterized by their total angular momentum,
$j_q=s_q+L$, where $s_q$ is the  light quark spin and $L$ is its 
orbital angular momentum.  $j_q$ is combined with $S_Q$, the spin of 
the heavy quark, to give the total angular momentum of the meson.  The 
quantum numbers $S_Q$ and $j_q$ are individually conserved. 
Thus, the four $L=1$ $P$-wave mesons can be grouped into two 
doublets characterized by the angular momentum of the light quark 
$j_q=3/2$ with $J^P=1^+, \; 2^+$ and $j_q=1/2$ with $J^P=0^+, \; 1^+$ 
where $J$ and $P$ are the 
total angular momentum and parity of the excited meson. 
In the heavy quark limit 
the members of the doublets will be degenerate in mass which is broken 
by $1/m_Q$ corrections \cite{isgur91,eichten93}.  
Similarly, heavy quark symmetry and the 
conservation of parity and $j_q$ predicts that the strong decays 
$D_{(s)J}^{(*)}(j_q=3/2)\to D^{(*)}\pi(K)$ proceed only through a 
$D$-wave while the decays 
$D_{(s)J}^{(*)}(j_q=1/2)\to D^{(*)}\pi(K)$ proceed only via an 
$S$-wave \cite{Godfrey:1986wj,Lu:1991px}.  
The states decaying to a $D$-wave are expected to be narrow
while whose decaying to an $S$-wave are expected to be broad. The 
properties of the charmed and charm-strange $j_q=3/2$ states 
are consistent with this prediction.

The observed properties of the $D_{sJ}^*(2317)$  \cite{Aubert:2003fg}
and %the subsequent observation of 
the $D_{sJ}(2460)$ \cite{Besson:2003cp}
did not agree with most theoretical predictions for the $j_q=1/2$ 
states. While most models predicted 
the $j=1/2$ doublet to be broad, decaying to $DK$ and $D^*K$ for the 
$J=0$ and $J=1$ state respectively,
the states observed by Babar and CLEO were 
below the $DK$ and $D^*K$ thresholds and
were quite narrow, decaying to $D^+_s\pi^0$ and $D_s^{*+}\pi^0$ 
respectively.
This led to considerable theoretical speculation about the 
nature of these states ranging from a conventional $c\bar{s}$ meson to 
exotic $DK$ molecules or tetraquarks \cite{Colangelo:2004vu}.  
A number of authors 
pointed out that if the masses of the conventional $c\bar{s}$ states 
are taken to be the measured $D_{sJ}^{(*)}$ masses, 
the predicted $D_{sJ}^{(*)}$ properties agree with the measured properties
\cite{Colangelo:2004vu,Bardeen:2003kt,Godfrey:2003kg,Colangelo:2003vg}.
While the $D_{sJ}^{(*)}$ 
states have attracted most of the attention, the first observations of the
the non-strange $j_q=1/2$ $P$-wave partner states, the $D_0^*$ and $D_1'$, 
\cite{Abe:2003zm,Link:2003bd,Anderson:1999wn}
has also taken place.  Comparing the theoretical predictions
for these $D_J^{(*)}$ states  
against their observed properties is an 
important complementary test of the reliability of the 
models \cite{Colangelo:2004vu,Godfrey:2004ct,Close:2005se}.  
This in turns acts as 
a baseline against which to measure the predicted properties of
the $D_{sJ}$ states. 

In this note we examine the properties of the charmed $P$-wave mesons.  
We concentrate on the nonstrange states, recalculating the strong 
widths using the measured masses and present new results on radiative 
transitions of these states. These are the main new results presented 
here \cite{Godfrey:2004ct}.  We begin with a summary of the 
experimental properties of charmed $P$-wave mesons including recent 
results.  This is followed by 
the new results on the properties of the charmed $P$-wave 
mesons.  We then briefly revisit the charm-strange $P$-wave mesons, 
discussing 
possible explanations of their unexpected properties.  
In the last section we conclude with some final comments and 
suggestions for further measurements.

\begin{table*}[t]
\caption{Summary of experimental measurements of the $L=1$ charmed 
mesons properties.} 
\vskip 0.5cm
\begin{tabular}{llrrrrrr} 
\hline
State & Property & PDG \cite{pdg}
		& Belle \cite{Abe:2003zm} 
		& FOCUS \cite{Link:2003bd}  
		& CLEO \cite{Anderson:1999wn} 
		& CDF \cite{Gorelov:2004ry}	
		& Average \\ \hline
%	&	& (MeV) & (MeV) & (MeV) & (MeV) & (MeV) & (MeV) \\
$D^{*0}_2$ & Mass (MeV) & $2458.9\pm 2$ & $2461.6\pm 3.9$ 
			& $2464.5 \pm 2.2$ &  
			& $2463.3\pm 1.0$ & $2462.7\pm 0.9$ \\
	& Width  (MeV) & $23\pm 5$ & $45.6\pm 8.0$ 
			& $38.7\pm 6.0$ & & $49.2\pm 2.4$ & $43.8\pm 2.0$ \\
$D^{*+}_2$ & Mass (MeV) & $2459 \pm 4$ &  
			& $2467.6 \pm 1.7$ & & & $2466.3\pm 1.6$  \\
	& Width (MeV) & $25^{+8}_{-7}$ & 
			& $34.1\pm 7.7$ & & & $29.4 \pm 5.4$ \\
$D_1^0$ 	& Mass (MeV) &  $2422.2\pm 1.8$ &  $2421.4\pm 1.7$
			& & & $2421.7 \pm 0.9$ & $2421.7\pm 0.7$ \\
 	& Width (MeV) & $18.9^{+4.6}_{-3.5}$  & $23.7\pm 4.8$
			& & & $20.0 \pm 2.1$ & $20.3\pm 1.7$ \\
$D_1'$  & Mass (MeV) & & $2427\pm 42$ 
			&  & $2461^{+53}_{-48}$ & & $2441\pm 32$ \\
	& Width (MeV) & & $384^{+130}_{-105}$ 
			& &$290^{+110}_{-91}$ & & $329 \pm 76$ \\
$D^*_0$ & Mass (MeV) & & $2308\pm 36$ & & & & $2308\pm 36$\\
	& Width (MeV) & & $276\pm 66$ & & & & $276\pm 66$ \\
\hline
\end{tabular}
\label{spectrum_table}
\end{table*}

\section{The Charmed P-wave Mesons}

In this section we study predictions for 
charmed $P$-wave meson properties and compare them to the experimental 
measurements.  The results for the 
strong decay widths are an update of previous results \cite{Godfrey:1986wj}
taking into account the measured masses. We also present the 
results of a new calculation of radiative transitions and suggest 
some tests of the models.

\subsection{Experimental Summary of charmed $P$-wave Mesons}

We start by summarizing the experimental measurements of
charmed $P$-wave meson properties in Table I.
Note that the recent  Belle \cite{Abe:2003zm}, 
FOCUS \cite{Link:2003bd}, and CDF \cite{Gorelov:2004ry}	measurements
of $\Gamma(D_2)$ % =37\pm 4.0$~MeV and $\Gamma(D_1^0)=23.7\pm 4.8$~MeV 
are larger than the PDG values.  Belle and FOCUS attribute 
differences with the older results to taking into account 
interference with the broader $D$ states.  
In the final column of Table I  the various experimental 
measurements are combined into weighted averages.    

In addition to the Belle and CLEO observations of the 
$D_0^*$ and $D_1'$ states
FOCUS \cite{Link:2003bd} reports two broad states, $D^{+}$ and $D^0$ with 
$M(D^+)=2403\pm 38$~MeV and $\Gamma(D^+)=283\pm 42$~MeV and 
$M(D^0)=2407\pm 41$~MeV and $\Gamma(D^0)=240\pm 81$~MeV.  They are unable to 
distinguish whether the broad states are due to the $D_0^*$ or $D_1'$
or whether both states contribute because of the large width.

\subsection{Spectroscopy}

Mass predictions are a first test of QCD motivated potential models 
\cite{Godfrey:1986wj,Godfrey:1985xj,zeng95,ebert97,de01,gj95,lahde00,dai}
and other calculational approaches 
\cite{Bardeen:2003kt,Kalashnikova:2001ig,lewis,cqm} 
used to confront the data. In 
QCD-motivated models the spin-dependent splittings test the Lorentz 
nature of the confining potential 
\cite{Godfrey:1986wj,Godfrey:2004ya,Blundell:1995au,ebert97}.  
Furthermore, the observation of 
heavy-light mesons is an important validation of heavy quark effective 
theory \cite{isgur91} and lattice QCD calculations \cite{lewis}. 
In Table II we summarize 
predictions for the charmed $P$-wave mesons.  The two $J=1$ states are 
linear combinations of $^3P_1$ and $^1P_1$ because for $q$ and 
$\bar{q}$ of different flavour,
charge conjugation parity, $C$, is no longer a good quantum number.  
We label these the $D_1$ and $D_1'$ defined as:
\begin{eqnarray}
D_1 & = & ^1P_1 \cos\theta +^3P_1 \sin\theta \nonumber \\
D_1' & = & -^1P_1 \sin\theta +^3P_1 \cos\theta 
\end{eqnarray}
with the mixing angle determined by the details of the model.  In the 
quark model it is due to the spin-dependent $LS$ mixing 
but more generally there are other contributions such  as
coupling via common decay channels.  If the mixing is dominated by 
decay channels the mixing might not be well represented by the orthogonal 
mixing given by Eqn. 1 \cite{bs05}. 
Measurement of the ratio of D/S decay 
amplitudes for the $D_1$ and $D_{1s}$ states could provide an 
important test of this possibility \cite{bs05}.
However, for the $^31P_1-^11P_1$ mixing, because
the states are almost degenerate it is 
likely that the linear combinations due to decay channel loops 
will be also be orthogonal \cite{barnes05}. 

Quark model calculations and heavy quark symmetry
predict that the 4 $L=1$ $c\bar{q}$ (where $q=u$ or $d$) are grouped 
into two doublets with properties characterized by the angular 
momentum of the lightest quark, $j_q=1/2$ and $j_q=3/2$. The heavy 
quark limit correponds 
to two physically independent mixing angles 
$\theta=-\tan^{-1}(\sqrt{2})\simeq -54.7^\circ$ and 
$\theta=\tan^{-1}(1/\sqrt{2})\simeq 35.3^\circ$ \cite{barnes}.
%$\sin\theta=\sqrt{2/3}$ and $\cos\theta=\sqrt{1/3}$. 
The $j_q=3/2$ states are identified with the previously observed 
$D_2^*(2460)$ 
and $D_1(2420)$ states while the $j=1/2$ have only recently been observed.  

\begin{table}[b]
\caption{Predictions for the masses of the charmed
$P$-wave mesons (rounded to the nearest MeV). The experimental masses 
are taken from Table I with errors only shown for the $D_0^*$ and 
$D_1'$ where the errors are relatively large.  The $J=1$ 
states are linear combinations of the $^3P_1$ and $^1P_1$ states.  The 
prime superscript is used for the state which in the HQL 
classification has $j_q=1/2$ and the unprimed state is used for the 
$j_q=3/2$ state. CM refers to chiral multiplets and 
LGT refers to the lattice gauge theory result.}
%\begin{center}
\begin{tabular}{l l l c c } \hline
Reference & $D_0^*(^3P_0)$ & $D_1'$  & $D_1$ & $D^*_2(^3P_2)$  \\ \hline
Experiment \footnotemark[1] & $2308\pm 36$ & $2441\pm 32$ 	
			& 2422  & 2463  \\
GI \cite{Godfrey:1985xj,Godfrey:1986wj} & 2399 	& 2460 	& 2470 	& 2502  \\
BEH [CM] \cite{Bardeen:2003kt} \footnotemark[1] & 2212 & 2355	&	&	\\ 
ZVR  \cite{zeng95} 	& 2270	& 2400	& 2410	& 2460  \\
EGF \cite{ebert97} 	& 2438	& 2501	& 2414	& 2459  \\
DE  \cite{de01} 	& 2377  & 2490 	& 2417  & 2460  \\
GJ  \cite{gj95} 	& 2279  & 2407  & 2421  & 2465  \\
LNR  \cite{lahde00} 	& 2341  & 2389  & 2407  & 2477  \\
DHJ \cite{dai}		& 2254	& 2391	& 2415	& 2461  \\
KNS \cite{Kalashnikova:2001ig} 
			& 2280 & 2354 & 2403 & 2432 \\
LW [LGT] \cite{lewis} \footnotemark[2]
			& 2444 	& 2413 	& 2422 	& 2453  \\
\hline
\end{tabular}
\footnotetext[1]{These values are for the neutral states.}
\footnotetext[2]{The lattice results are taken by adding the results 
of Ref. \cite{lewis} to the mass of the $D^0$ state.
The $^3P_1$ and $^1P_1$ masses are given in the $D_1$ and $D_1'$ 
columns respectively. Errors for the lattice results are not shown.}
%\end{center}
\end{table}

Predictions from various quark model and other calculations are 
summarized and compared to the experimental masses in Table II.
There is a considerable spread amoung the predictions.  
The mass splittings of the GI model
\cite{Godfrey:1986wj,Godfrey:1985xj} are in good
agreement with the measured splittings although the $P$-wave c.o.g. is 
$\sim 40$~MeV too high.  Other models are consistent with the two 
previously observed states but do not do particularly well with one or 
more of the $j_q=1/2$ states.  In any case, the spread of predictions 
underlines the importance of more precise measurements to test these 
models.

\subsection{Strong Transitions}

The $P$-wave meson strong decays can be described by $D$ and $S$-wave 
amplitudes 
\cite{Rosner:1985dx,Godfrey:1985xj,Godfrey:1986wj,Lu:1991px,Close:2005se}.  
The amplitude formulas for the decays are given in Table 
III.  To obtain 
values for $D$ and $S$ we rely on models of meson decay.  Here, we 
give results using the pseudoscalar emission model
and update the  predictions of Godfrey and 
Kokoski \cite{Godfrey:1986wj}
by adjusting the phase space for the $D_0^*$ and $D_1'$ decays.
The $^3P_0$ \cite{Close:2005se}, 
flux-tube models \cite{Godfrey:1986wj} and chiral 
quark model calculations \cite{de01} give 
qualitatively similar results.
In the pseudoscalar emission model the $D$ and $S$ amplitudes are 
given by \cite{Godfrey:1985xj,Godfrey:1986wj}:
\begin{equation}
D=A_Q \left( { m_Q \over {m_Q+m_{\bar{q}}}} \right)
\left( { q\over \beta } \right)^2
\left( { \beta \over \beta_Q} \right) 
\left( {q \over {2\pi} } \right)^{1/2}
F(q^2)
\end{equation}
and 
\begin{equation}
S=S_Q \left( {q \over {2\pi} } \right)^{1/2} F(q^2)
\end{equation}
where
\begin{equation}
F(q^2)=\exp \left[ { - 
\left( { m_Q \over {m_Q+m_{\bar{q}}}} \right)^2 {q^2\over{4\beta_Q^2}}
} \right]
\end{equation}
and $m_q=m_u=m_d=0.3$~GeV, $m_c=1.7$~GeV are the relevant constituent 
quark masses used in the decay calculation, 
$\beta=0.4$~GeV and $\beta_Q=0.5$ are harmonic oscillator wavefunction 
parameters used in obtaining these amplitudes.  $\beta$ is 
taken from the light meson decay analysis of Ref. \cite{Godfrey:1985xj} and 
$\beta_Q$ was obtained by fitting the rms radii of HO wavefunctions to the 
rms radii of the GI wavefunctions.  $S_Q=3.27$ and $A_Q=1.67$ come 
from the light meson decay analysis of Ref. \cite{Godfrey:1985xj}.  
In Table III
$\theta_0 =\tan^{-1}(\sqrt{1/2})$ arises
from the Clebsch-Gordan coefficients of the $^3P_1$ and $^1P_1$
contributions to the decay amplitudes. 
We stress that the 
calculated widths are predictions of the model with no free parameters.
The results are compared to experimental measurements in Table III. 
The $D_0^*$ and $D_1'$ masses used in Table III came about from 
adjusting the $P$-wave c.o.g. of the GI calculation downward by 40~MeV 
so that 
the $D_2^*$ and $D_1$ masses are in better agreement with experiment
in order to give more reliable phase space estimates for the $D_0^*$ and 
$D_1'$ decays.

Overall the agreement between theory and experiment is good.  
The experimental widths for the $D_2^*$ and $D_1$ are the weighted 
averages given in Table I obtained by averaging the PDG values 
\cite{pdg} and the Belle \cite{Abe:2003zm}, FOCUS \cite{Link:2003bd}  and 
CDF \cite{Gorelov:2004ry} results.  
As already noted the Belle and FOCUS widths 
are larger than the PDG values and are in better agreement with our 
results than the PDG values.  
However, it should be noted that there is some sensitivity in these 
results to phase space.  
(The observant reader has probably noted 
that we included the $D_2^{*0}$ rather than the smaller (but with 
larger error) $D_2^{*+}$ width.  Nevertheless, we consider the results 
perfectly acceptable given the limitations of the model \cite{bg}.)

Treating
the $D$ and $S$ amplitudes ({\it modulo} the $q^{2L+1}$ phase space factor)
and mixing angle as free parameters does not qualitatively improve the 
agreement with experiment.  Similarly, fitting the mixing angle to the 
$D_1$ and $D_1'$ widths results in a mixing angle consistent with both the 
GI result and the HQL within errors. 

In the HQL the $D_1$ decay is 
purely $D$-wave and the $D_1'$ decay is purely $S$-wave.  Because the 
$S$-wave partial width is so large, even relatively small deviations 
from the HQL would broaden the $D_1$ width quite substantially.  Thus, 
these strong decays are a good test of the HQL. 

Considering the experimental uncertainties and the inherent 
limitations of the simple decay model the agreement is excellent.  It 
would be interesting to see if the agreement survives comparison with 
future, more precise measurements.

\begin{table*}
\caption{Predictions for strong decay widths of the charmed $P$-wave 
mesons.  All widths are given in MeV.
The $D_2^*$ and $D_1$ masses are taken from Table I and the masses 
used for the $D_1'$ and $D_0^*$ are described in the text.  We took
$m_\pi=140$~MeV/c$^2$. 
}
%\begin{center}
\begin{ruledtabular}
\begin{tabular}{l l c c c c c } %\hline
Decay & Amplitude Formula & $M_i$ & $M_f$ & q & Width  & Expt. Width \\
%	&		& (MeV) & (MeV) & (MeV) & (MeV) & (MeV)  \\
\hline 
$D_2^* \to D^*\pi$ & $ -\sqrt{3\over{10}} D$ 
		& 2463 & 2010 & 391 & 18 & \\
$\qquad \to D\pi$ & $ -\sqrt{1\over{5}} D$ 
		& 2463 & 1869 & 507 & 37 & \\
$\qquad \to D^*\pi + D\pi$ & & & &  & 55 & $43.8\pm 2.0$ \\
$D_1 \to [D^*\pi]_S$ & $ \sqrt{1\over{2}} \sin(\theta+\theta_0) S$ 
		& 2422 & 2010 & 354 & 7 & \\
$\qquad \to [D^*\pi]_D$ & $ \sqrt{1\over{2}} \cos(\theta+\theta_0) D$ 
		& 2422 & 2010 & 354 & 18 & \\
$\qquad \to [D^*\pi]_S + [D^*\pi]_D$ & & & &  & 25 & $20.3\pm 1.7$ \\
$D_1' \to [D^*\pi]_S$ & $ \sqrt{1\over{2}} \cos(\theta+\theta_0) S$ 
		& 2420 & 2010 & 352 & 244 & \\
$\qquad \to [D^*\pi]_D$ & $ -\sqrt{1\over{2}} \sin(\theta+\theta_0) D$ 
		& 2420 & 2010 & 352 & 0.5 & \\
$\qquad \to [D^*\pi]_S + [D^*\pi]_D$ & & & &  & 244 & $329\pm 76$ \\
$D_0^* \to D\pi$ & $ -\sqrt{1\over{2}} S$ 
		& 2359 & 1869 & 421 & 277 & $276\pm 66$  \\
\end{tabular}
\end{ruledtabular}
\end{table*}

\subsection{Electromagnetic Transitions}

Radiative transitions probe the internal structure of 
hadrons
\cite{Godfrey:2003kg,Bardeen:2003kt,Colangelo:2003vg,Colangelo:2005hv} 
and as such give an additional tool to understand hadronic states.  Because 
the recently observed $D_1'$ and $D_0^*$ states are quite broad it is 
unlikely that radiative decays of these states will be observable.  
However,  it is {\it a priori} possible that they might be observed for the 
$D_2^*$ and $D_1$ states.  The decays $D_1\to D^*\gamma$ and 
$D_1\to D\gamma$  would be especially interesting as they would give 
some insights into the $^3P_1-^1P_1$ mixing. 

The $E1$ radiative transitions are given by \cite{Kwo88}
\begin{eqnarray} 
\Gamma(i & \to &  f + \gamma)  \\
& = & \frac{4}{27} \alpha \; 
\langle e_Q \rangle^2 \; 
\omega^3 \;(2J_f +1) \; |\langle ^{2s+1}S_{J'} | r | ^{2s+1}P_J \rangle |^2 
\;{\cal S}_{if} \nonumber
\end{eqnarray}
where ${\cal S}_{if}$ is a statistical factor with ${\cal S}_{if}=1 $
for the transitions between spin-triplet states
($D_{J}^{(*)}(1P) \to D^*\gamma$)
and ${\cal S}_{if}=3$  for the transition between spin-singlet states
($D_{1}\to D\gamma$),
$\langle e_Q \rangle$ is an effective quark charge given 
by 
\begin{equation}
\langle e_Q \rangle = {{m_q e_c - m_c e_{\bar{q}}}\over{m_c+m_q}}
\end{equation}
where $e_c= 2/3$ is the charge of the $c$-quark
and $e_{\bar{d}}=1/3$, $e_{\bar{u}}=-2/3$ are the charges of the  
$d$ and $u$ antiquarks given in units of $|e|$, $m_c=1.628$~GeV, 
$m_q=0.22$~GeV are the mass of the $c$ and $q=u$, $d$ quarks taken from Ref. 
\cite{Godfrey:1985xj},  $\alpha = 1/137.036$ is the fine-structure constant,
and $\omega$ is the 
photon's energy.  The matrix elements $\langle S | r | P \rangle$ 
given in Table IV were evaluated using the wavefunctions of 
Ref. \cite{Godfrey:1985xj}.
Relativistic corrections are included in the E1 transition 
via Siegert's theorem \cite{siegert,mcclary,moxhay} 
by including spin dependent interactions in the Hamiltonian used to 
calculate the meson masses and wavefunctions.   To calculate the 
appropriate photon energies the PDG \cite{pdg} values were used for 
observed mesons while the predictions from Ref. \cite{Godfrey:1985xj} 
were used 
for unobserved states with the following modification:  While
splittings between $c\bar{q}$ states predicted by Ref. \cite{Godfrey:1985xj}
are in good agreement with experiment, the masses are 
slightly higher than observed so to give a more reliable 
estimate of phase space, the masses used in Table IV have been adjusted 
down by 40~MeV from the predictions of Ref. \cite{Godfrey:1985xj}.

A final subtlety is that 
the $J=1$ states are linear combinations of $^3P_1$ and $^1P_1$ 
as described by eqn. 1.  The radiative widths were caculated using
$\theta=-26^o$ %and the conventions of Ref. \cite{Godfrey:1986wj}
and include the appropriate factors $\cos^2\theta$ and $\sin^2\theta$.

Table IV gives the quark model predictions for 
E1 radiative transitions between the $1P$ and $1S$ charmed 
mesons \cite{Godfrey:2004ct}.  One should appreciate that the 
predictions for the BR's are imprecise given the uncertainties in the 
strong widths used to calculate the BR's.
For completeness 
we include results for both the charged and neutral $D$ states but due 
to cancellations in $\langle e_Q \rangle$ 
the radiative widths of the charged states are much 
smaller than those of the neutral states. The radiative decays of the 
narrow states are the most likely to be observed.
The $D_1^0\to 
D^{*0}\gamma$ and $D_1^0\to D^{0}\gamma$ transitions are of particular 
interest since the ratio of these partial widths are a measure of the 
$^3P_1-^1P_1$ mixing angle in the charmed meson sector:
\begin{equation}
{{\Gamma(D_1\to ^3S_1 + \gamma)}\over{\Gamma(D_1\to ^1S_0 + 
\gamma)}} = {{\omega_t^3 |\langle r \rangle_t|^2}\over
{\omega_s^3 |\langle r \rangle_s|^2}}
{{\sin^2\theta}\over{\cos^2\theta}}
\end{equation}
and can therefore test how well the HQL is satisfied.  As already 
mentioned, measurement of the $^3P_1-^1P_1$ mixing angle could reveal 
mixing effects due to decay channel coupling \cite{bs05} 
which would shed light on the nature of the new $D_{sJ}^{(*)}$ states.

\begin{table*}
\caption{Partial widths and branching ratios for 
E1 transitions between $1P$ and $1S$ charmed mesons.  
The $M_i$ and the total widths used to calculate the BR's 
are taken from Table I.  The matrix elements are calculated using the 
wavefunctions of Ref. \cite{Godfrey:1985xj}. To calculate the BR's we 
used the following total widths: 
for the $D_{2}^{*+}$ and $D_{2}^{*0}$ the averages given in Table I,
for $D_{1}^+$ and $D_{1}^0$ the average for the $D_1^0$ given in Table I,
and for the $D_1'$ and $D_0^*$ the predicted widths given in Table III. 
}
\begin{center}
\begin{ruledtabular}
{\begin{tabular}{@{}l l c c c c c c@{}} 
Initial & Final & $M_i$ & $M_f$ &  $k$ & 
	$\langle 1P | r | nS \rangle $ &  Width  & BR  \\
state  & state & (MeV) & (MeV) & (MeV) & (GeV$^{-1}$) & (keV) & \\
\hline
$D_{2}^{*+}$ & $D^{*+} \gamma $ & 2466 & 2010 & 414 & 2.367 & 59 
		& $2.0\times 10^{-3}$ \\
$D_{2}^{*0}$ & $D^{*0} \gamma $ & 2463 & 2007 & 414 & 2.367 & 572 
		& $1.3\times 10^{-2}$ \\
$D_{1}^+$    & $D^{*+} \gamma$ & 2422 & 2010 & 377 & 2.367 & 8.6 
		& $4.2\times 10^{-4}$ \\
	   & $D^{+} \gamma$ & 2422 & 1869 & 490 & 2.028 & 58 
		& $2.9\times 10^{-3}$ \\
$D_{1}^0$    & $D^{*0} \gamma$ & 2422 & 2007 & 379 & 2.367 & 85 
		& $4.2\times 10^{-3}$ \\
	   & $D^{0} \gamma$ & 2422 & 1865 & 493 & 2.028 & 574 
		& $2.8\times 10^{-2}$ \\
$D_{1}'^+$    & $D^{*+} \gamma$ & 2420 & 2010 & 375 & 2.367 & 36 
		& $1.4\times 10^{-4}$ \\
	   & $D^{+} \gamma$ & 2420 & 1869 & 488 & 2.028 & 14 
		& $5.6\times 10^{-5}$ \\
$D_{1}'^0$    & $D^{*0} \gamma$ & 2420 & 2007 & 378 & 2.367 & 352 
		& $1.4\times 10^{-3}$ \\
	   & $D^{0} \gamma$ & 2420 & 1865 & 491 & 2.028 & 135 
		& $5.5\times 10^{-4}$ \\
$D_{0}^{*+}$ & $ D^{*+} \gamma$ & 2359 & 2010 & 323 & 2.345 & 28 
		& $1.0\times 10^{-4}$ \\
$D_{0}^{*0}$ & $ D^{*0} \gamma$ & 2359 & 2007 & 326 & 2.345 & 274 
		& $9.9\times 10^{-4}$ \\
%\hline
\end{tabular}}
\end{ruledtabular}
\end{center}
\end{table*}

\subsection{Discussion}

Overall the agreement between quark model predictions and 
experiment is quite good.  
Although the GI mass prediction for the $P$-wave c.o.g. is
slightly high,
the splittings are in good agreement with experiment.  The strong 
decay widths also agree well with experiment.  In fact, the predicted 
widths agree better with the more 
recent Belle, FOCUS, and CDF measurements than the older PDG values.  More 
precise measurements would be welcomed to further test the models.
Note that the physical $D_1^{(\prime)}$ 
states are linear combinations of the $^3P_1$ and $^1P_1$ states so 
that the good agreement for the decay widths reflects a successful 
prediction for the $1^3P_1 -1^1P_1$ mixing angle.  This can be further 
tested by measuring the $D_1^0\to D^{*0}\gamma$ and $D_1^0\to D^{0}\gamma$
partial widths.

The overall conclusion is that the $P$-wave charmed mesons 
are well described by the quark model and   
models invoked to describe the $D_{sJ}^*(2317)$ and  
$D_{sJ}(2460)$ states must also explain their non-strange charmed meson 
partners.

\section{The Charm-Strange $P$-wave Mesons}

Motivated by the successful description of the charmed $P$-wave mesons we 
briefly revisit the charm-strange $P$-wave mesons.  We start by 
comparing in Table V 
the observed properties to the quark model predictions of 
Ref. \cite{Godfrey:2003kg} and \cite{Godfrey:1986wj}.
%predicted properties of these states 
The predicted properties are shown assuming the measured 
masses of the $D_{sj}(2317)$ and $D_{sj}(2460)$ states.  
The predictions using the quark model mass predictions can be found in
Ref. \cite{Godfrey:2003kg}.

The narrow $j=3/2$ states are identified with the $D_{s1}(2536)$ 
and $D_{s2}(2573)$ states.  Their observed properties are in good 
agreement with quark model predictions \cite{Godfrey:1986wj,Godfrey:1985xj}.
In contrast, the $j=1/2$ states were predicted 
to be broad and to decay to $DK$ and $D^*K$ and had not been 
previously observed.  
The recently discovered $D^*_{sJ}(2317)$ is below $DK$ threshold and the 
$D_{sJ}(2460)$ is below $D^*K$ threshold so the only allowed strong 
decays are $D_{sJ}^{(*)}\to D_s^{(*)}\pi^0$ which violates isospin and is 
expected to have a small 
width \cite{Godfrey:2003kg,Bardeen:2003kt,Colangelo:2003vg}.  This led 
to considerable speculation about the nature of these states 
\cite{Colangelo:2004vu}.  
However, if one assumes they are the conventional $j_q=1/2$ 
$c\bar{s}$ states,  {\it albeat} with a much lower mass than generally 
expected, their properties can be calculated using models of hadrons.

The strong decays  $D_{sJ}^{(*)}\to D_s^{(*)}\pi^0$ and radiative
transitions were calculated by a number of authors
\cite{Godfrey:2003kg,Bardeen:2003kt,Colangelo:2003vg}.  
They concluded that radiative transitions should have large BR's and 
are important diagnostic probes for understanding the nature of 
these states \cite{Godfrey:2003kg,Bardeen:2003kt,Colangelo:2003vg}.  

Although there are discrepancies between some of the 
quark model predictions and existing measurements they can easily be
accomodated by the uncertainty in theoretical estimates of 
$\Gamma(D_{sJ}^{(*)}\to D_{s}^{(*)}\pi^0)$ 
and by adjusting the $^3P_1-^1P_1$ mixing angle 
for the $D_{s1}$ states. 
For example, Ref. \cite{Bardeen:2003kt} predicts a 
$\Gamma(D_{sJ}^{(*)}\to D_{s}^{(*)}\pi^0)$ width about twice as large 
as the values used to calculate the BR's in Table V which were taken from 
Ref. \cite{Godfrey:2003kg}.  One should also note that there is still 
considerable uncertainty in the experimental width measurements.
As in the case of the $D_1$ states, the 
radiative transitions to $D_{s}$ and $D^*_s$ can be used to constrain
the $^3P_1-^1P_1$ ($c\bar{s}$) mixing angle using eqn. 7.  

The problem with the newly found $D_{sJ}$ states are the mass 
predictions. Once the masses are fixed the narrow widths follow.  
As a first step to understanding the discrepancy between quark model 
predictions and the observed masses
we revisited the relativized quark model \cite{Godfrey:1985xj}
to see if the observed masses 
could be accomodated with a change of the model's parameters. We were 
not able to find a set of parameters that could accommodate the masses of 
the new states while at the same time preserving the successful mass 
predictions of the model.  

A possible solution, long suggested in the literature, is that 
the strong $S$-wave coupling of the $D_{sJ}^{(*)}$ states
to the  $DK$ ($D^*K$) decay channel
(and the nearness to the $D^{(*)}K$ thresholds) shifts the respective 
masses 
\cite{vanBeveren:2003kd,Rupp:2004rf,Hwang:2004cd,Simonov:2004ar,Becirevic:2004uv,eichten2004}.  
Including these coupled channel effects,
van Beveren Rupp and Kleefeld \cite{vanBeveren:2003kd,Rupp:2004rf}, 
Hwang and Kim \cite{Hwang:2004cd}, Simonov and Tjon \cite{Simonov:2004ar}, 
and Becirevic Fajfer and Prelowsek \cite{Becirevic:2004uv} are able to 
explain the low $D_{sJ}^*(2317)$ and $D_{sJ}(2460)$ masses.
However,  others found that when more and more intermediate states are 
included the mass of the state is not stable \cite{bs05}. 
It was also found that coupled channel effects appear to lead to 
comparable shifts in states that were previously in good agreement 
with experiment \cite{swanson}.  
It was suggested that including these 
virtual meson loops has, to a large extent,
the effect of ``renormalizing'' the string 
tension except near thresholds %(and spin dependent effects)
\cite{Geiger:1989yc}.
The exception to this result is for 
the low lying $0^{++}$ states where Geiger and Isgur 
found that loop effects manifest themselves as large shifts in the 
masses \cite{Geiger:1992va}.  Isgur suggested that this was due to 
S-wave channels having a cusp discontinuity at threshold 
\cite{Isgur:1998kr}.  This special behavior
of the $0^{++}$ channel was also noted by Kalashnikova
\cite{Kalashnikova:2005ui}.  The resulting 
mass shifts depend on the position 
of the valence mass relative to threshold.  For the $q\bar{q}$ 
$D_{s0}^{(*)}$ mass prediction relative to the $S$-wave $DK$ threshold 
ref. \cite{Isgur:1998kr} expects a large negative shift 
\cite{Isgur:1999cd}.  
We should stress, however, that other calculations applied to scalar 
meson masses find that the shifts 
are rather modest, of order tens of MeV \cite{VanBeveren:1986ea} and 
that the loop effects result in weak binding of the pseudoscalars
\cite{vanBeveren:2001kf}.  This is van Beveren and Rupp's 
explanation of the $D_{sJ}^{(*)}$
states \cite{vanBeveren:2003kd}.
One concludes that one way or another 
the strong $S$-wave coupling to the 
$D_{sJ}^{(*)}$ states  with a nearby threshold  
is the likely solution to the puzzle, but it is not clear what the 
exact mechanism is, nor has this been 
unquestionably demonstrated.  To do so one would have to demonstrate 
not only that the contributions to the $D_{sJ}$ states converge, but 
that one can successfully explain all states including those that are 
already well described by the constituent quark model. In other words, 
states that are well described remain so and the agreement of others 
improves.

\begin{table*}
\caption{Comparison of predicted and measured charm-strange $P$-wave meson 
properties.  The BR's are based on the theoretical values for the 
partial widths.
Experimental numbers come from the PDG \cite{pdg}. The theoretical 
predictions come from Ref. \cite{Godfrey:2003kg} 
except for the strong decay widths of the 
$D_{s2}^*$ which comes from Ref. \cite{Godfrey:1986wj} rescaled for 
the correct phase space.}
%\begin{center}
\begin{ruledtabular}
\begin{tabular}{l l r r c } %\hline
State & Property & Expt. & Theory &  BR  \\
\hline 
$D_{s2}^*$ & Mass (MeV) & $2573.5\pm 1.7$  & 2574 & \\
	& $\Gamma(D_{s2}^*\to D^*K)$ (MeV) & & 1  & \\
	& $\Gamma(D_{s2}^*\to DK)$ (MeV) & & 20  & \\
	& $\Gamma(D_{s2}^*\to D_s^* \gamma )$ (keV) 
			&  & 19 & $\sim 1.3 \times 10^{-3}$\\
	& $\Gamma_{Total}$ (MeV) & $15^{+5}_{-4}$  & 21  & \\
$D_{s1}$ & Mass (MeV) & $2535.3\pm0.31$  & 2535 & \\
	& $ \Gamma(D_{s1}\to D^*K)$ (keV) & & 340\footnotemark[1] & 97 \%  \\
	& $ \Gamma(D_{s1}\to D_s^* \gamma )$ (keV) &  & 5.6 & 1.6\%   \\
	& $ \Gamma(D_{s1}\to D_s \gamma )$(keV) &  &  15  &	4.2\%\\
	& $ \Gamma_{Total}$ (keV) & $<2300$ (90\% CL)  & 371  & \\
$D_{s1}'(2.463)$ & Mass (MeV) & $2458.9\pm 0.9$ &  & \\
	& $ \Gamma(D_{s1}'\to D_s^* \gamma )$ (keV) &  & 5.5 &  24\% \\
	& $\Gamma(D_{s1}'\to D_s^+ \gamma )$ (keV) &  & 6.2 &	27\% \\
	& $\Gamma(D_{s1}'\to D_s \pi^0 )$ (keV) &  & $\sim 10 $	& 43\% \\
	& $\Gamma(D_{s1}'\to D_s \pi\pi )$ (keV) & & $\sim 1.6$ & 7\% \\
	&$ \frac{\Gamma(D_s^{*+} \gamma )}{\Gamma( D_s^{*+} \pi^0 )}$ 
		& $<0.16$ (90\% CL) & 0.55  &  \\
	&$ \frac{\Gamma(D_s^{+} \gamma )}{\Gamma( D_s^{*+} \pi^0 )}$ 
		& $0.31\pm 0.06$  & 0.62  &  \\
	&$ \frac{\Gamma(D_s^{+} \pi^+\pi^- )}{\Gamma( D_s^{*+} \pi^0 )}$ 
		& $0.14\pm 0.045$  & 0.16  &  \\
$D_{s0}^*(2317)^+$ & Mass (MeV) & $2317.3\pm 0.6$ & & \\
	& $ \Gamma(D_{s0}^{*+} \to D_s^* \gamma )$ (keV) 
			& & 1.9 & $\sim 16$ \%  \\
	& $ \Gamma(D_{s0}^{*+} \to D_s \pi^0 )$ (keV) 
			& &  $\sim$ 10 & $\sim 84\%$  \\
	& $ \frac{\Gamma(D_s^{+} \gamma )}{\Gamma(D_s^{+} \pi^0 )}$ 
			& $<0.05$ (90\% CL)  & 0  &  \\
	& $ \frac{\Gamma(D_s^{*+} \gamma )}{\Gamma(D_s^{+} \pi^0 )}$ 
			& $<0.059$ (90\% CL)  & $\sim 0.19$  &  \\
\end{tabular}
\end{ruledtabular}
%\end{center}
\footnotetext[1]{The PDG gives $\Gamma < 2.3$~MeV 90\% C.L.. 
We used the width given in Ref. \cite{Godfrey:1986wj} rescaled for phase space.}
\end{table*}

\section{Summary}

To summarize, we found that the charmed $P$-wave  mesons are well 
described by the quark model.  However, it is important to confirm the 
broad $j=1/2$ states and obtain more precise measurements of their 
properties. In particular, measuring the radiative decay BR's of the 
$D_1(2420)$ measures the $^3P_1-^1P_1$ mixing angle and is a good test 
of the HQL.  In contrast, 
the $D^*_{sJ}(2317)$ and $D_{sJ}(2460)$ states have 
masses lower than expected for the missing $0^+$ and $1^+$ $j=1/2$ 
$c\bar{s}$ states.  
We suggest that the strong $S$-wave coupling of the 
$j_q=1/2$ $c\bar{s}$ states to $DK$ (and $D^*K$) 
is the key to the unusual properties of the new light $D_{sJ}$ mesons
but further work is needed for a definitive answer.  
Radiative transitions are important diagnostic tests 
of the nature of these states and should be pursued.

\acknowledgments

The author thanks Ted Barnes for helpful communications and 
the TRIUMF theory group for their hospitality where 
some of this work took place.
This research was supported in part by the Natural Sciences and Engineering 
Research Council of Canada.

%\begin{references}

%\end{references}

\end{document}